\documentclass[aps,showpacs,twocolumn,prl]{revtex4}

\IfFileExists{srcltx.sty}{\usepackage[active]{srcltx}}%

\usepackage{epsfig}
\usepackage{graphicx}
\usepackage{epsfig}
\usepackage{bm}
 
\newcommand{\beq}{\begin{equation}}
\newcommand{\eeq}{\end{equation}}
\newcommand{\bea}{\begin{eqnarray}}
\newcommand{\eea}{\end{eqnarray}}

\def\simle{\lower 2pt \hbox {$\buildrel < \over {\scriptstyle \sim }$}}
\def\simge{\lower 2pt \hbox {$\buildrel > \over {\scriptstyle \sim }$}}

\begin{document}

\preprint{UCLA/05/TEP/35}

\title{Relic keV sterile neutrinos and reionization}


\author{Peter L. Biermann$^{1,2,3}$ and Alexander Kusenko$^4$
}

\affiliation{$^1$Max-Planck Institut for Radioastronomy, Bonn, Germany\\
$^2$Department of Physics and Astronomy, University of Bonn, Germany\\ 
$^3$Department of Physics and Astronomy, University of Alabama, Tuscaloosa, 
AL 35487, USA \\
$^4$Department of Physics and Astronomy, University of California, Los
Angeles, CA 90095-1547, USA }


\begin{abstract}

A sterile neutrino with mass of several keV can account for cosmological
dark matter, as well as explain the observed velocities of pulsars. 
We show that X-rays produced by the decays of these relic sterile neutrinos
can boost the production of molecular hydrogen, which can speed up the
cooling of gas and the early star formation, which can, in turn, lead to a
reionization of the universe at a high enough redshift to be consistent with
the WMAP results.  

\end{abstract}

\pacs{14.60.St, 95.35.+d,   \hfill UCLA/05/TEP/35}

\maketitle

\renewcommand{\thefootnote}{\arabic{footnote}}
\setcounter{footnote}{0}

There is an ample evidence that most of the matter in the universe is not
ordinary atomic matter, but that it is made up of new, yet undiscovered
particles.   A seemingly unrelated astrophysical puzzle is the origin of the
rapid motions of pulsars, whose velocities range in hundreds of km/s, while as
many as 15\% of pulsars have speeds in excess of
1000~km/s~\cite{Kusenko:review}. It is intriguing that these two puzzles may
have a simultaneous solution if there exists a sterile neutrino with mass of
several keV and a small mixing with ordinary neutrinos.  If such a particle
exists, it would be produced in the early universe with the right abundance to
be the dark matter~\cite{dm_s,Fuller,Abazajian:2005gj}, and it would 
have an important effect on the cosmological stucture
formation~\cite{Abazajian:2005xn}. The same particle would be emitted from a
supernova with a sufficient anisotropy to give the pulsar a kick velocity of
the right magnitude~\cite{Kusenko:review,ks97,fkmp}. The pulsar kick from an
anisotropic emission of sterile neutrinos can enhance the convection during the
first second of the supernova, which can increase the energy of the supernova
shock and bring the supernova calculations in better agreement with
observations~\cite{Fryer:2005sz}. Another hint at the existence of such a
particle comes from the observations~\cite{Fan01} of central galactic black
holes with masses $3 \times 10^6$ to $3 \times 10^9$ solar masses at redshifts
as high as 6.4.  Dark matter in the form of sterile neutrinos could speed up
the formation of supermassive black holes at high redshift \cite{MB05}.
Theoretical considerations lead one to believe that at least three sterile
neutrinos should exist below the electroweak scale~\cite{Asaka:2005an_three},
in which case the neutrino oscillations can explain the baryon asymmetry of the
universe~\cite{Akhmedov:1998qx,Asaka:2005pn_baryons}. 

Structure formation and reionization of the universe can be used to test some
of the properties of dark-matter particles.  Here we show that X-ray photons
from the decays of the relic sterile neutrinos could help speed up the
reionization of the intergalactic medium (IGM), in accordance with the WMAP
observations of the early ionization at redshift $z_{\rm r}=17\pm
5$~\cite{WMAP-polar,WMAP}.

If the intergalactic medium (IGM) was reionized by stars (rather than some
new physics~\cite{Hansen:2003yj}), the WMAP data imply a
very early and efficient star formation. Sterile neutrinos 
with keV masses have a non-negligible free-streaming length, which could delay
the small-scale structure formation~\cite{Yoshida:2003rm}, if it were not 
for the photons emitted in their decays.  The width of the decay is extremely
small, and the sterile neutrinos are stable on cosmological time scales. 
Nevertheless, some of them do decay producing photons with
keV energies. If the dark matter is made up of sterile neutrinos, the X-ray
photons from their decay can catalyze the formation of molecular hydrogen in
the amounts that can speed up the gas cooling and star formation, hence 
leading to an early reionization.  

The decays of particles during "dark ages" and their impact on
reionization were discussed in Refs.~\cite{Chen:2003gz,Mapelli:2005hq}.
According to Mapelli and Ferrara~\cite{Mapelli:2005hq}, the
electron fraction $x_e$ remains between 0.001 and 0.01 at redshifts
$10<z<100$. These values are too small to produce the appreciable
Thompson optical depth, but they are considerably higher than the electron
fraction in the absence of sterile neutrinos.  Mapelli and
Ferrara~\cite{Mapelli:2005hq} concluded that the X-ray photons from sterile
neutrino decays can play no role in reionization.  However, even a small
fraction of ions may be sufficient to catalyze the formation of $H_2$, which,
in turn, can precipitate a rapid star formation.  Then the early stars can
reionize the IGM.  We will show that the electron fraction as high as $10^{-3}$
can, in fact, boost the production of molecular hydrogen, $H_2$, to a value
which is more than sufficient for a rapid star formation well before $z_{\rm
r}=17\pm 5$.

Let us consider a singlet neutrino that has a non-zero mixing with the electron
neutrino, and let us ignore other mixings for simplicity.  Then the mass
eigenstates have a simple expression in terms of the weak eigenstates:
\begin{eqnarray}
| \nu_1 \rangle & = & \cos \theta \, | \nu_e \rangle - \sin \theta  \, |
\nu_s  \rangle \\
| \nu_2 \rangle & = & \sin \theta \, | \nu_e \rangle + \cos \theta \, |
\nu_s \rangle  .
\label{eigenstates}
\end{eqnarray}

If the mixing angle $\theta $ is small, one of the mass eigenstates,
$\nu_1$ behaves very much like a pure $\nu_e$, while the other, $\nu_2$, is
practically ``sterile'', which means it has weak interactions suppressed by
a factor $(\sin^2 \theta)$ in the cross section. The mixing enables 
the mass
eigenstate $\nu_2$ to decay into lighter neutrinos, as well as $\nu_1$ and a
photon.  The inverse width of $\nu_2 \rightarrow \nu_1 \gamma$ decay is 
\beq
\tau
= 1.3 \times 10^{26} {\rm s} \left ( \frac{7 \ {\rm keV}}{m_s} \right )^5
 \left ( \frac{ 0.8\times 10^{-9}}{\sin^2 \theta} \right ).
\eeq
The X-ray photon produced in this two-body decay has energy $E_\gamma
= m_s/2$.  

The mass range consistent with $\nu_2$ being dark matter, with the
pulsar kicks~\cite{Kusenko:review}, and all the other
constraints 
\cite{Mapelli:2005hq,Fuller,x-rays,Abazajian:2005gj,MB05,Abazajian:2005xn,
Kusenko:2004qc} is
$$2 \, {\rm keV} < m_s < 8\,{\rm keV}. $$
The corresponding photon energies are
$$1\, {\rm keV} < E_\gamma < 4\, {\rm keV}.
$$
This range is rather conservative.   A broader range of parameters may be
allowed, depending on cosmology~\cite{low_reheat}.

The attenuation length $\lambda$ for real absorption of such photons in
hydrogen~\cite{PDG} is $ 0.12 \, {\rm g/cm}^2 < \lambda < 7.7 \, {\rm g/cm}^2
$. The dominant absorption process is the ionization edge of hydrogen
\cite{RB79}. Since the density of gas at redshift
$z$ is $\rho_{0,H} =4\times 10^{-31} (1+z)^3 {\rm g \, cm^{-3}},$ the energy
attenuation length of a photons becomes smaller than the horizon size $ 
H^{-1} = H_0^{-1} (1+z)^{-3/2}$ at redshift $z> z_{\rm thick} = 9$ for
$E_\gamma = 1 \, {\rm keV}$, and at $z> z_{\rm thick}=148$ for $E_\gamma = 4 \,
{\rm keV}$.  This includes a correction for the presence of helium.  However,
even at lower redshifts, $z<z_{\rm thick}$, the photons can ionize gas after
they get redshifted. As the photon gets red-shifted, the cross section goes up
with $(1+z)^{-3}$ due to the ionization edge frequency behavior. The mean free 
path decreases as $(1+z)^{-3/2}$.  Hence, the photons emitted at redshift $z_a$
get all absorbed at some lower redshift $z_b$.  For 4~keV photons, 
$7.7 ((1+z_b)/(1+z_a))^3 \; = \; 0.0043 (1+z_b)^{3/2}$. Photons
emitted at redshift 77 are all absorbed at redshift 40, implying a loss 
of about factor 2 in the energy available for ionization and heating.  Lower
redshifts of the emission yield even greater losses in energy that
can be used for ionization.  One can safely assume that all the photons
produced at redshift greater than $z \, \simge \,  40$ lose most of their energy
to heating and ionization of the intergalactic medium within less than the
Hubble time.

Interactions of $1-4$~keV photons with gas start out by ionization which
absorbs the photon and ejects an energetic electron, which heats and
ionizes further.  Compton  scattering transfers only a small fraction of 
the photon energy to the electrons.   A cascade started by a single photon can
ionize a number of atoms.  Photoionization and Compton ionization of atoms are
the dominant energy loss processes~\cite{x-ray_abs}.  The photon energy is
divided between reionization, heating, and excitations, with about 
30\% of it going to ionization~\cite{x-ray_Shull}.  Hence the number of
ionized atoms can be estimated as 
$\sim \eta (m_s /13.6 {\rm eV}) $ per X-ray photon, where $\eta \approx 
0.3 $.

\begin{figure}[t]
\centering
\epsfxsize=8cm \epsfbox{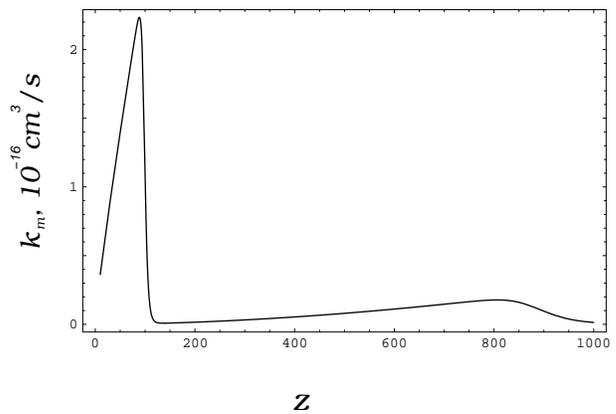}\\[3mm]
\caption[fig1]{\label{fig1}
The rate $k_{m}$ of molecular hydrogen production as a function of redshift.
For illustrative purposes, the gas temperature is assumed to be close to that
of the CMBR (which is a good assumption, at least, for $z>150$). 
}                
\end{figure}

The density of photons produced from the sterile neutrino decays at time $t$ 
is the sterile neutrino density times $(t/\tau)=(t_0/\tau) (1+z)^{-3/2} $. 
Each of these photons then ionizes $\sim \eta (m_s /13.6 {\rm eV}) $ atoms. In
the absence of other effects, at redshift $z \, \simge \, 40$, the fraction of
ionized atoms is
\beq
x_e^{(s)}\sim \frac{0.2}{(1+z)^{3/2}}
\left ( \frac{\eta}{0.3}
\right )
\left
(\frac{m_s}{7 \ {\rm keV}} \right )^5
\left ( \frac{\sin^2\theta}{ 0.8\times 10^{-9}} \right ).
\eeq
Thus, the production of $H^+$ ions due to sterile neutrino decays occurs 
at a constant rate per atom,
\beq
A =  1.4  \times  10^{-16} \, {\rm s}^{-1}  
\left (\frac{m_s}{7 \ {\rm keV}} \right )^5
\left ( \frac{\sin^2\theta}{ 0.8\times 10^{-9}} \right ).
\label{A}
\eeq
The free electrons are consumed in the reaction $H^+e^-
\rightarrow H \gamma$ with the rate $k_1\approx 1.88 \times 10^{-10}
(T_{_K})^{-0.64}
{\rm cm}^3 {\rm s}^{-1}$, where $T_{K}$ is the gas temperature in Kelvins.
Hence, the electron fraction is described by the equation
\beq
\dot{x_e} = A- k_1 n_{H} x_e^2 = A - B \left ( \frac{t_0}{t} \right)^{1.6}
x_e^2,
\label{eqx}
\eeq
where $n_{H}= n_{0,H} (1+z)^3$, $n_{0,H} =2 \times 10^{-7} {\rm 
cm}^{-3} $,
and $B=k_1 \, n_{0,H}=2 \times 10^{-17} {\rm s^{-1}}$.

The star formation depends on cooling, which becomes very efficient
when molecular hydrogen is produced.   The critical $H_2$ fraction
needed for a rapid star formation is $f_c \approx 5\times
10^{-4}$~\cite{Tegmark:1996yt}.

\begin{figure}[t]
\centering
\epsfxsize=8cm \epsfbox{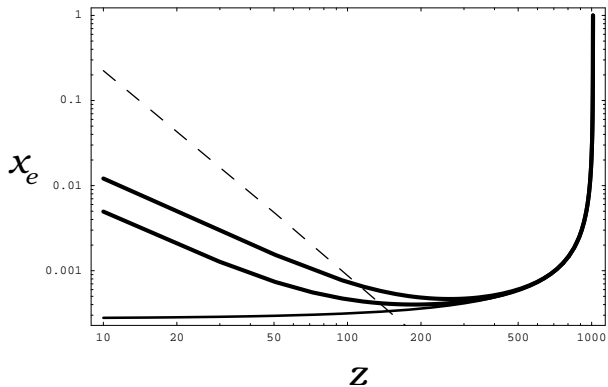}\\[3mm]
\caption[fig2]{\label{fig2}
The fraction of ions, $x_e$, in the absence of sterile neutrinos (thin 
line),
and for the dark-matter sterile neutrinos with masses 4 and 7 keV (lower and
upper thick lines, respectively). Also shown is the limit of baryon 
decoupling from CMBR (dashed line).  Below the dashed line, gas cooling is 
unaffected by the CMBR.  
}                
\end{figure}

In the presence of ions, the molecular hydrogen can form in several
reactions. One channel involves $H^-$   
\bea
H+e^- & \rightarrow & H^-+ \gamma \\
H^-+H & \rightarrow & H_2+e^-
\eea
and proceeds at a high rate for $z\ll 200$.  However, since $H^-$ is quickly
destroyed at high temperature by the interactions with the cosmic microwave
background (CMB) photons, this channel is closed at higher redshifts.
At $z\gg 200$, the main reactions are
\bea
H^{+}+H & \rightarrow & H_2^++ \gamma, \\
H_2^{+}+H & \rightarrow & H_2+H^+.
\eea
Both channels depend on the presence of ions and free electrons, and, hence,
on the value of $x_e$.  The fraction of molecular hydrogen can be found 
from
the following equation
\bea
\dot{f} & = & k_m(t)\, n_{H}(t)\, x_e (t) \ \left (1-x_e(t)-2 f(t) \right )
\nonumber \\
& \approx & k_m(t)\, n_{H}(t)\, x_e (t),
\label{eqf}
\eea
where $k_m$ is a function defined in Ref.~\cite{Tegmark:1996yt}.  The behavior
of $k_m$ as a function of redshift is shown in Figure 1. It reaches a maximum
$k_{m, {\rm max}} \approx 2 \times 10^{-16} {\rm cm}^3 {\rm s}^{-1}$ at
$z\approx 80$.  In the relevant range of parameters, the molecular hydrogen is
not destroyed~\cite{GP98}, and the right-hand side of eqn.~(\ref{eqf}) remains
positive.

Shortly after recombination, the ionization fraction $x_e \sim 1$, and
some molecular hydrogen is produced.  However, in
the absence of sterile neutrinos, $x_e$ falls below $10^{-3}$ by redshift
$z=400$, at which point further production of molecular hydrogen is stymied by
the deficit of free electrons.  In our case, the free electrons are
continuously supplied by the decays of $\nu_s$ at the rate given by
equation~(\ref{A}).

We can now integrate equation (\ref{eqx}) with the initial condition $x_e=1$
at recombination.  The result can be used to obtain $f(t)$ from equation
(\ref{eqf}).  Figure~\ref{fig2} shows the solution of equation 
(\ref{eqx}) for three values of $A$ corresponding to dark-matter sterile
neutrino masses 4 and 7 keV, as well as for $A=0$.  

Let us now estimate the fraction of molecular hydrogen.  Since $H_2$ is
produced and not destroyed for $z \ll 10^3$, the value of $f(z)$ is 
always greater than the fraction produced at any earlier time.  In particular,
$f(z)$ is greater than the fraction of hydrogen generated during one Hubble
time at redshift $z_1$, as long as $z_1 > z$. In other words,
\bea
f(z) & > &  \Delta_z f(t_1) \approx \frac{df}{dt} \times t_1 \nonumber \\
& \approx &  k_m(z_1)\, n_{_H}(z_1) \, x_e(z_1) \, \frac{t_0}{(1+z_1)^{3/2}}.
\label{delta_f}
\eea

Let us consider a gas cloud that has virialized at some redshift $z_{\rm
vir}$. The evolution of this cloud takes place as usual~\cite{Tegmark:1996yt},
as long as the electron fraction is not high enough to maintain the thermal
equilibrium between the gas and the CMBR at redshift close to  $z_{\rm 
vir}$. However, if the gas interacts strongly with the CMBR, the Compton
scattering keeps its temperature close to that of the CMBR, and the analyses of
Ref.~\cite{Tegmark:1996yt} do not apply. Compton cooling is described by the
following equation:
\beq
\frac{dT}{dt}= (T_\gamma - T) \, k_{\rm compt} \ x_e ,
\eeq
where $k_{\rm compt}=2.6 \times 10^{-20} (1+z)^4\, {\rm s}^{-1}$. To check the
efficiency of the Compton cooling, one must compare the cooling rate 
with the
expansion rate of the universe.  This gives a limit on the fraction of 
ions.
The Compton cooling is inefficient (and can be neglected) as long as
\beq
x_e < 0.9 \times 10^{-3} \left ( \frac{100}{1+z} \right )^{5/2}.
\eeq
This limit is shown in Figure \ref{fig2} as a dashed line.  As one can see,
for the parameters of interest, the decoupling of gas from the CMBR occurs at
$z>100$. For smaller redshifts, one can use the semi-analytical results 
from
Ref.~\cite{Tegmark:1996yt}, making a correction for a higher electron
fraction.

One should also check that the gas temperature is not raised 
significantly by
the presence of keV photons.  If the gas temperature reaches $3000$~K, the
molecular hydrogen is destroyed. The total energy released in photons
from the sterile neutrino decays is
\bea
L & = & n_{\nu_s} (m_s/2) \tau^{-1}  \nonumber \\
& = & 8 \times 10^{-30} \frac{\rm erg}{ \rm cm^3 \, s}\left ( \frac{1+z}{100}
\right)^3 \nonumber \\
& & \times
\left (\frac{m_s}{7 \ {\rm keV}} \right )^5
\left ( \frac{\sin^2\theta}{ 0.8\times 10^{-9}} \right ).
\eea
One can compare this heating process with the cooling processes discussed in
Ref.~\cite{Tegmark:1996yt}.  For the relevant range of parameters, the heat
deposited by sterile neutrinos can be neglected.

Let us now consider a gas cloud collapsing and virializing at some redshift
$z_{\rm vir}$.  The density in a recently virialized cloud is about a
factor $18 \pi^2 $ higher than the mean hydrogen density~\cite{Tegmark:1996yt}:
$$
n_{\rm vir}= 23 \, {\rm cm}^{-3} \left ( \frac{1+z}{100}
\right)^3.
$$
Let us consider $z_{\rm vir} =100$.  The fraction of molecular hydrogen
produced at this redshift, according to eqn.~(\ref{delta_f}), is
\bea
f & > &  \Delta_{100} f  =  9 \times 10^{-4} \nonumber \\
& \times & \left ( \frac{k_m}{10^{-16}\, {\rm cm}^3\, s^{-1}} \right)  
\left ( \frac{x}{ 10^{-3}} \right)  
\left ( \frac{n_{\rm vir}}{20 \, {\rm cm}^{-3} } \right ).
\eea

This lower bound on the molecular hydrogen fraction is higher than the
critical value $f_c\approx 5 \times 10^{-4}$ derived by Tegmark {\em et al.}
for a successful collapse, especially near the maximum of $ k_m $. 
This maximum is achieved at $z=80$, as shown
in Figure~\ref{fig1}, and one expects the most efficient production of $H_2$
for $z_{\rm vir} \sim 10^2$.  In the absence of sterile neutrinos, the
electron fraction is around $3 \times 10^{-4}$, for which $f $ drops below
$f_c$.  

In conclusion, if sterile neutrinos with masses of several keV 
constitute the dark matter, their decays produce enough photons to boost the
production of molecular hydrogen in clouds collapsing at redshifts as
high as $z\sim 100$. The presence of molecular hydrogen facilitates rapid
cooling and the early star formation.  At the same time, the sterile neutrinos
can have a non-negligible free-streaming length, which can affect the early 
halo collapse~\cite{dm_s,Abazajian:2005xn}. The full analysis must combine the
enhanced production of molecular hydrogen with simulations of dark matter halo
formation.  Such detailed analysis is beyond the scope of this {\em Letter}; it
will be presented elsewhere.  

The authors thank F.~Munyaneza, M.~Shaposhnikov, J.~Stasielak, and E.~Wright
for very helpful comments. The authors thank the Aspen Center for Physics for
hospitality.  The work of P.L.B. was supported by the Pierre
Auger grant 05~CU~5PD1/2 via DESY/BMBF. The work of A.K. was supported in part
by the DOE grant DE-FG03-91ER40662 and by the NASA ATP grants NAG~5-10842 and
NAG~5-13399.

\end{document}